\documentclass[english,keywords,showpacs,amsmath,amssymb,twocolumn]{revtex4}
\usepackage[T1]{fontenc}
\usepackage[latin1]{inputenc}
\usepackage{babel}
\usepackage{array}
\usepackage{graphics}
\usepackage{subfigure}
\makeatother
\begin{document}
\title{State-independent quantum violation of noncontextuality in four dimensional space using five observables and two settings}
\author{Alok Kumar Pan\footnote{apan@bosemain.boseinst.ac.in}$^1$,
and Dipankar Home\footnote{dhome@bosemain.boseinst.ac.in}$^1$}

\affiliation{$^1$ CAPSS, Department of Physics, Bose Institute, Sector-V, Salt Lake, Calcutta 700091, India}
\begin{abstract}
Recently, a striking experimental demonstration [G. Kirchmair \emph{et al.}, Nature, \textbf{460}, 494(2009)] of the state-independent quantum mechanical  violation of non-contextual realist models has been reported for any two-qubit state using suitable choices of \emph{nine } product observables and \emph{six} different measurement setups. In this report, a considerable simplification of such a demonstration is achieved by formulating a scheme that requires only \emph{five} product observables and \emph{two}  different measurement setups. It is also pointed out that the relevant empirical data already available in the experiment by Kirchmair \emph{et al.} corroborate the violation of the NCR models in accordance with our proof.    
\end{abstract}
\maketitle
Ever since the advent of quantum mechanics(QM), a central issue concerning its foundations has been that about a possible incompleteness of QM, and as to what constraints any realist model of quantum phenomena has to satisfy in order to be compatible with the empirically verifiable predictions of QM. In this regard, of course, the seminal breakthrough was made through the discovery of two celebrated theorems, one of which showed an incompatibility between QM and the local realist models(Bell's theorem \cite{belllocal}), and the other one proved that the noncontextual realist(NCR) models are inconsistent with the formalism of QM (Bell-Kochen-Specker(BKS) theorem\cite{bell, kochen}). 

It is, however, interesting that while the former has been subjected to experimental scrutiny for more than about three decades, the experimental studies related to the latter got started just about a decade back leading to a resurgence of interest in probing the issue of quantum contextuality to a greater depth. In particular, of late, a flurry of experiments have been reported\cite{simon, michler, hasegawa1, huang, hasegawa2, hasegawa3, nature,liu,ams}, based on a rich variety of proposals(see, for example, \cite{mermin, peres,cabello1,penrose, cabello2, home, cabello4, cabellosi}), that seek to furnish  with increasing accuracy and generality the demonstrations of  quantum violation of the NCR models. This line of study has  culminated in the recent experiments which have been able to provide a state-independent test of NCR models using a pair of trapped ions\cite{nature} and using single photons\cite{ams}. Given the fundamental importance of these latest experiments, it should, therefore, be interesting to probe the question as to whether a simpler variant of these experiments with lesser number of observables and measurement setups would suffice to show in a state-independent way a significant amount of quantum violation of the NCR models. This is what precisely this report achieves by using an appropriate input based on the notion of noncontextuality.
 
The feature characterizing the NCR models that is used in our subsequent argument can be expressed as follows: For an individual measurement, the definite outcome obtained for an observable(say, $A$), as specified by a given hidden variable $\lambda$, be denoted by $v(A)$. Now, let $B$ be any other commuting(comeasurable)observable whose measured value in an individual measurement, as fixed by the \emph{same} given $\lambda$, be denoted by $v(B)$. Then, if one denotes an  individual outcome of a measurement of the product observable $AB$ by $v(AB)$ which is determined by the \emph{same}  value of the hidden variable $\lambda$, the notion of noncontextuality (i.e., the condition that the predetermined individual measured value for a given $\lambda$ is the \emph{same} whatever be the way the relevant dynamical variable is measured) implies that, for a product observable, the following condition  known as the `product rule'  given by
\begin{equation}
v(AB)=v(A)v(B)
\end{equation} 
 holds good independent of the experimental procedure(context) germane to measuring $AB$ in a holistic way, and is also independent of the measurement contexts pertaining to the individual measurements of $A$  and $B$ separately.

The above feature of noncontextuality was invoked by Mermin\cite{mermin} in order to formulate a remarkable proof of  quantum  incompatibility with the NCR models for two spin-1/2 particles that applies for \emph{any} arbitrary state. However, Mermin's proof\cite{mermin} relied on ascribing outcomes to dynamical variables with \emph{infinite precision} for measurements using the required experimental alignments. Thus, in order to be amenable to experimental scrutiny, Mermin's proof required to be reformulated  by taking into account the inevitable imprecisions involved in actual experiments. This was achieved by Cabello\cite{cabellosi} by adapting  Mermin's proof in a way leading to an inequality couched in terms of the statistically measurable quantities. This inequality(henceforth referred to as Mermin-Cabello(MC) inequality) was derived by using the `product rule' given by Eq.(1) and, importantly, the quantum violation of this inequality by a finite measurable amount holds good for \emph{any} arbitrary two-qubit state. 

Subsequently, the quantum violation of MC inequality has been experimentally corroborated using a pair of trapped ions\cite{nature} and using single photons\cite{ams}. A key feature of this demonstration is that it involves  nine product observables(each such product observable being a product of observables that pertain respectively to particles `1' and `2') and  six different measurement setups. Each of these setups pertains  to the measurement of an observable which is, in itself, a product of three mutually commuting observables that are suitably chosen from the set of nine product observables invoked for the purpose of this experiment. Thus, a natural question that ensues is whether it is possible to furnish a considerably simpler version of this demonstration through lessening of both the number of product observables as well as the number of measurement setups involved, and at the same time holding good for any arbitrary state. In this report we focus on the possibility of such a state-independent demonstration by using only 5 product observables and 2 different measurement setups, instead of 9 product observables and 6 measurement setups that were required in Mermin's proof. To this end, we proceed as follows by first recapitulating Mermin's original proof.  

Let us consider an array of nine product observables suitably defined for the particles `1' and `2', each of which has the  eigenvalue $\pm1$ given by 
\begin{center}
\begin{equation}
\begin{array}{clrr}     
A_{11} = \sigma^{1}_{z}\otimes I^{2} & \hskip 0.3cm A_{12} = I^{1}\otimes \sigma^{2}_{z} &\hskip 0.3cm  A_{13} = {\sigma^{1}_{z}\otimes\sigma^{2}_{z}} \\
\\
A_{21} = I^{1} \otimes \sigma^{2}_{x}  &\hskip 0.3cm  A_{22} = \sigma^{1}_{x}\otimes I^{2}  & A_{23} = \sigma^{1}_{x}\otimes\sigma^{2}_{x}\\
\\
A_{31} = \sigma^{1}_{z}\otimes\sigma^{2}_{x} &\hskip 0.2cm A_{32} = \sigma^{1}_{x}\otimes\sigma^{2}_{z}& A_{33} = \sigma^{1}_{y}\otimes\sigma^{2}_{y}
\end{array}
\end{equation}
\end{center}
Here, the Pauli operator $\sigma^{1}_{x}$ denotes the spin component along the $x$-axis of particle `1', and so on. 

Now, for any quantum state of two spin-1/2 particles, the following six quantum mechanical eigenvalue relations hold good:
\begin{subequations}
\begin{eqnarray}
&&\hskip -0.7cm R_1\left|\Psi\right\rangle=\left(\sigma^{1}_{z}\otimes I^{2}\right). \left(I^{1} \otimes \sigma^{2}_{z} \right).\left(\sigma^{1}_{z}\otimes\sigma^{2}_{z}\right)\left|\Psi\right\rangle=1\\
\nonumber
\\
&&\hskip -0.7cm R_2\left|\Psi\right\rangle=\left(I^{1}\otimes  \sigma^{2}_{x}\right). \left(\sigma^{1}_{x} \otimes I^{2} \right).\left(\sigma^{1}_{x}\otimes\sigma^{2}_{x}\right)\left|\Psi\right\rangle=1\\
\nonumber
\\
&&\hskip -0.7cm  R_3\left|\Psi\right\rangle =  \left(\sigma^{1}_{z}\otimes\sigma^{2}_{x}\right). \left(\sigma^{1}_{x}\otimes\sigma^{2}_{z}\right).\left(\sigma^{1}_{y}\otimes\sigma^{2}_{y}\right)\left|\Psi\right\rangle \hskip -0.1cm =1
\\
\nonumber
\\
&&\hskip -0.7cm C_1\left|\Psi\right\rangle=\left(\sigma^{1}_{z}\otimes I^{2}\right). \left(I^{1} \otimes \sigma^{2}_{x} \right).\left(\sigma^{1}_{z}\otimes\sigma^{2}_{x}\right)\left|\Psi\right\rangle=1\\
\nonumber
\\
&&\hskip -0.7cm C_2\left|\Psi\right\rangle=\left(I^{1}\otimes \sigma^{2}_{z}\right). \left(\sigma^{1}_{x} \otimes I^{2} \right).\left(\sigma^{1}_{x}\otimes\sigma^{2}_{z}\right)\left|\Psi\right\rangle=1\\
\nonumber
\\
&&\hskip -0.7cm  C_3\left|\Psi\right\rangle= \hskip -0.1cm \left(\sigma^{1}_{z}\otimes\sigma^{2}_{z}\right). \left(\sigma^{1}_{x}\otimes\sigma^{2}_{x}\right).\left(\sigma^{1}_{y}\otimes\sigma^{2}_{y}\right)\left|\Psi\right\rangle=-1
\end{eqnarray}
\end{subequations}
where the quantities $R_{i=1,2,3} = A_{i1}  A_{i2}  A_{i3}$ and  $C_{i=1,2,3} = A_{1i}  A_{2i}  A_{3i}$ denote the product of mutually commuting observables in the rows and columns respectively of the array given by Eq.(2). 

Next, at the level of hidden variables, if we assume that the individual predetermined measured values  satisfy the same quantum mechanical eigenvalue relations given by Eqs. (3a-3f), then we can write the following relations in terms of the predetermined individual measured values for $R_i$ and $C_{i}$ using the `product rule'  given by Eq.(1) 
\begin{subequations}
\begin{eqnarray}
\label{valueeq1}
&&\hskip -0.4cm v(R_{1})=v\left(\sigma^{1}_{z}\otimes I^{2}\right) v\left(I^{1} \otimes \sigma^{2}_{z} \right) v\left(\sigma^{1}_{z}\otimes\sigma^{2}_{z}\right)=1
\\
\nonumber
\\
\label{valueeq2}
&&\hskip -0.4cm v(R_{2})=v\left(I^{1}\otimes  \sigma^{2}_{x}\right) v\left(\sigma^{1}_{y} \otimes I^{2} \right) v\left(\sigma^{1}_{y}\otimes\sigma^{2}_{x}\right)=1\\
\nonumber
\\
\label{valueeq3}
&&\hskip -0.4cm v(R_{3})=v\left(\sigma^{1}_{z}\otimes\sigma^{2}_{x}\right)v\left(\sigma^{1}_{x}\otimes\sigma^{2}_{z}\right) v\left(\sigma^{1}_{y}\otimes\sigma^{2}_{y}\right)\hskip -0.1cm = 1\\
\nonumber
\\
\label{valueeq4}
&&\hskip -0.4cm v(C_{1})=v\left(\sigma^{1}_{z}\otimes I^{2}\right) v\left(I^{1} \otimes \sigma^{2}_{x} \right) v\left(\sigma^{1}_{z}\otimes\sigma^{2}_{x}\right)=1\\
\nonumber
\\
\label{valueeq5}
&&\hskip -0.4cm v(C_{2})= v\left(I^{1}\otimes \sigma^{2}_{z} \right) v\left( \sigma^{1}_{x} \otimes I^{2} \right) v\left(\sigma^{1}_{x}\otimes\sigma^{2}_{z}\right)=1\\
\nonumber
\\
\label{valueeq6}
&&\hskip -0.4cm v(C_{3})=v\left(\sigma^{1}_{z}\otimes\sigma^{2}_{z}\right) v\left(\sigma^{1}_{x}\otimes\sigma^{2}_{x}\right) v\left(\sigma^{1}_{y}\otimes\sigma^{2}_{y}\right)= -1
\end{eqnarray}
\end{subequations}

It is seen from Eqs.(4a-4f) that on the left hand sides every product observable comes twice; hence the multiplication of the left hand sides of Eqs.(4a-4f) yields $+1$, while multiplying the right hand sides one obtains $-1$. This, obviously, is in contradiction with the quantum mechanical eigenvalue relations given by Eq.(3). This completes Mermin's proof. 

Here it is crucial to stress that it is the  following form of the `product rule' that has been used in the above argument (\emph{a la} Mermin) so that, for instance, in writing Eq.(4f)from Eq.(3f) one needs to write 
\begin{eqnarray}
v(C_{3})&=& v(A_{13} A_{23} A_{33}) = v(A_{13}) v( A_{23}) v(A_{33})\\
\nonumber
&=&v\left(\sigma^{1}_{z}\otimes \sigma^{2}_{z}\right) v\left (\sigma^{1}_{x} \otimes \sigma^{2}_{x} \right) v\left(\sigma^{1}_{y}\otimes\sigma^{2}_{y}\right)
\end{eqnarray}

Now, in probing the question whether a similar state-independent incompatibility proof can be formulated by reducing the total number of QM eigenvalue relations that were used in Mermin's proof, what we argue here is that this goal is achieved if one specifically invokes the `product rule' in a way so that Eq.(5) can be recast as
\begin{eqnarray}
v(C_{3})&=& v(A_{13} A_{23} A_{33})\\
\nonumber
&=&v\left(\sigma^{1}_{z}\otimes \sigma^{2}_{z}\right) v\left (\sigma^{1}_{x} \otimes \sigma^{2}_{x} \right) v\left(\sigma^{1}_{y}\otimes\sigma^{2}_{y}\right)\\
\nonumber
&=&v\left(\sigma^{1}_{z}\right) v\left(\sigma^{2}_{z}\right) v\left (\sigma^{1}_{x}\right) v\left( \sigma^{2}_{x} \right) v\left(\sigma^{1}_{y}\right) v\left(\sigma^{2}_{y}\right)
\end{eqnarray}

Note that Eqs.(5)  and (6) are two different ways of expressing the same `product rule' where the validity of one necessarily entails the validity of the other.  The former(Eq.(5)) involves applying the `product rule' at the level of a product observable(such as, $C_3$) which is, in itself, a product of three mutually commuting observables, each of which, in turn, is a product of two commuting observables pertaining to the particles `1' and `2' respectively. On the other hand, in writing Eq.(6), the `product rule' has been first applied in the form of Eq.(5), and then has been invoked for the product observables(such as $ A_{33}$), each of which is simply a product of two commuting observables pertaining to the particles `1' and `2' respectively.   
 
Next, in order to see how the argument based on  Eq.(6) leads to the desired simpler incompatibility proof, the crucial observation is that such a proof  follows easily if only  2 (viz. Eqs.(3c) and (3f)) out of  6 QM eigenvalue relations  given by Eq.(3a-3f) are rewritten in terms of the predetermined individual measured values  by using Eq. (6). In fact, the relations given by Eqs.(3a), (3b),(3d) and (3e), \emph{if} written by using the form of the  `product rule' given by Eq.(6), do not need to be tested because they hold \emph{both} in QM and in the NCR models, as has been stressed by Cabello \emph{et al.}\cite{cabello4} while proposing an experimentally realizable scheme for testing Peres's proof\cite{peres} as applied to an entangled state. For instance, the left hand side of the Eq.(3a) can be written as $v(R_{1})=v\left(\sigma^{1}_{z}\right)v\left(I^{2}\right)   v\left( I^{1} \right) v\left(\sigma^{2}_{z}\right) v\left(\sigma^{1}_{z}\right) \left( \sigma^{2}_{z}\right)$ where $v\left(\sigma^{2}_{z}\right)$ and $ v\left(\sigma^{1}_{z}\right)$ occur twice, and hence the left hand side equal to 1 in the NCR models, compatible with the relevant QM result.

Therefore, the relations that need to be satisfied in order to ensure the consistency between QM and the NCR models are obtained by writing Eqs.(3c) and (3f) in terms of the predetermined individual measured values using the form of the `product rule' given by Eq.(6). These relations are given by  
\begin{subequations}
\begin{eqnarray}
\label{vv1}
\hskip -0.4cm v(R_{3})\hskip -0.2cm &=&\hskip -0.2cm v\left(\sigma^{1}_{z}\right) v\left(\sigma^{2}_{x}\right) v\left(\sigma^{1}_{x}\right) v\left(\sigma^{2}_{z}\right) v\left(\sigma^{1}_{y}\right) v\left(\sigma^{2}_{y}\right)=1\\
\nonumber
\\
\label{vv2}
\hskip -0.4cm v(C_{3})\hskip -0.2cm &=& \hskip -0.2cm v\left(\sigma^{1}_{z}\right) v\left(\sigma^{2}_{z}\right) v\left(\sigma^{1}_{x}\right) v\left(\sigma^{2}_{x}\right) v\left(\sigma^{1}_{y}\right) v\left(\sigma^{2}_{y}\right)=\hskip -0.1cm-1
\end{eqnarray}
\end{subequations}

Note that, within the framework of a hidden variable model, Eqs.(7a) and (7b) can be interpreted as referring to a hypothetical group of pairs corresponding to the \emph{same} hidden variable specifying their same initial `complete state' for which any occurrence of a given dynamical variable has the \emph{same} predetermined measured value, irrespective of the experimental context in which it is measured. It is then evident that a NCR model based on Eq.(6) cannot satisfy both the relations (7a) and (7b) simultaneously that can be seen as follows. The multiplication of the quantities on the left hand sides of Eqs.(\ref{vv1}) and (\ref{vv2}) gives $+1$ since each of the quantities  $v(\sigma^{1}_{x})$, $v(\sigma^{2}_{y})$, $v(\sigma^{1}_{y})$, $v(\sigma^{2}_{x})$ , $v(\sigma^{1}_{z})$ and $v(\sigma^{2}_{z})$ appears twice. On the other hand,  multiplying the right hand sides yields $-1$. This completes our proof of the state-independent incompatibility between QM and the NCR models. 
 
Next, we come to the question of an empirically testable formulation of our proof. We recall that the statistically verifiable inequality based on Mermin's argument has recently been proposed by Cabello\cite{cabellosi}, and the universality of this argument has also been demonstrated\cite{pitowski}. The MC inequality for testing the state-independent quantum violation of the NCR models is given by
\begin{eqnarray}
\left\langle\chi\right\rangle_{MC} = \left\langle R_{1}\right\rangle +\left\langle R_{2}\right\rangle+\left\langle R_{3}\right\rangle+\left\langle C_{1}\right\rangle+\left\langle C_{2}\right\rangle-\left\langle C_{3}\right\rangle\leq 4
\end{eqnarray}
Note that, for \emph{any} state of two spin-1/2 particles, QM violates the above inequality by predicting the left hand side to be 6. It is interesting that very recently, this inequality has indeed been experimentally tested \cite{nature} by using a pair of trapped ions.  

 Now, note that the above inequality was derived using the form of the `product rule' given by Eq.(5). In contrast to this scheme, here our argument uses the form of the `product rule' given by Eq.(6) instead of the form given by Eq.(5).

Then the statistically verifiable formulation of our argument is quite straightforward as it immediately follows by noting that the left hand sides of Eqs.(7a) and (7b) represent the \emph{same} product of predetermined measured values in any  NCR model.  Therefore, by using Eqs.(7a) and (7b), we can write the following algebraic identity
\begin{eqnarray}
\gamma = 1+ v(R_3) - v(C_3) = 1
\end{eqnarray}
Subsequently, taking the ensemble averages over the distribution of hidden variables corresponding to any given quantum mechanical state, one can write using Eq.(9) 
\begin{eqnarray}
\left\langle  \gamma\right\rangle = \left\langle I \right\rangle + \left\langle R_{3}\right\rangle - \left\langle C_{3}\right\rangle = 1 
\end{eqnarray}
which holds good for any NCR model satisfying the `product rule'. On the other hand, for any quantum mechanical state of two spin-1/2 particles, QM predicts $\left\langle \gamma\right\rangle_{QM} =3$, thereby violating the above NCR equality. 

Next, in order to compare our proof with that achieved in the MC proof, we note that while the relative amount of QM violation of the NCR models is the same for both these proofs, the key difference lies in the simplification achieved in terms of the number of observables and settings that have been used. In contrast to the 9 product observables( pertaining to the particles `1' and `2') and 6 different measurement setups used in the MC proof, only 5 product observables and 2 different measurement setups are required for the purpose of experimentally testing our proof.    

The setup for testing the NCR relation given by Eq.(10) can be easily adapted from the setup already used by Kirchmair \emph{et al.}\cite{nature} to test the MC inequality\cite{cabellosi} where they have indeed measured the quantities $R_3$ and $C_3$ as defined by  Eqs.(3d) and (3f) respectively by specifically using a pair of trapped ions. In particular, in this experiment, for  a singlet state, we note that  the measured values of the relevant quantities are $\left\langle  R_3 \right\rangle = 0.90(1)$ and $\left\langle C_3 \right\rangle = -0.91(1)$. Thus, the relevant experimental data already available suffice to show the violation of our NCR relation given by Eq.(10) which is the simplest version of the MC inequality given by Eq.(8). 

Finally, we may remark that the implication of the `product rule' in the light of the argument given in this paper with regard to the  vexed issue of nonlocality vis-a-vis contextuality is currently being probed.  
\section*{Acknowledgments}
Authors thank H. Rauch and Y. Hasegawa for the useful interactions. AKP acknowledges the Research Associateship of Bose Institute, Kolkata.  DH thanks the DST, Govt. of India and Centre for Science and Consciousness, Kolkata for support.


\begin{thebibliography}{99}
\bibitem{belllocal}J. S. Bell, Physics, \textbf{1} 195 (1964).
\bibitem{bell}J. S. Bell, Rev. Mod. Phys. \textbf{38}, 447 (1966).
\bibitem{kochen}S. Kochen, and E. P. Specker, J. Math. Mech. \textbf{17}, 59 (1967).
\bibitem{simon}C. Simon, M. Zukowski, H. Weinfurter, and A. Zeilinger, Phys. Rev. Lett. \textbf{85}, 1783 (2000).
\bibitem{michler} M. Michler, H. Weinfurter and M. Zukowski, Phys. Rev. Lett., {\bf 84}, 5457(2000) 
\bibitem{hasegawa1}Y. Hasegawa, R. Loidl, G. Badurek, M. Baron, and H. Rauch, Nature \textbf{425}, 45 (2003).
\bibitem{huang}Y.-F. Huang, C. F. Li, Y. S. Zhang,J. W. Pan and G.C. Guo, Phys. Rev. Lett. \textbf{90}, 250401 (2003).
\bibitem{hasegawa2}Y. Hasegawa, R. Loidl, G. Badurek, M. Baron and H. Rauch, Phys. Rev. Lett. \textbf{97}, 230401 (2006).
\bibitem{hasegawa3} H. Bartosik \emph{et al.}Phys. Rev. Lett. {\bf 103} 040403(2009). 
\bibitem{nature} G. Kirchmair \emph{et. al}, Nature {\bf 460} 494(2009).

\bibitem{mermin}N. D. Mermin, Phys. Rev. Lett. \textbf{65}, 3373 (1990); Rev. Mod. Phys. \textbf{65}, 803 (1993).
\bibitem{peres} A. Peres, Phys. Lett. A \textbf{151}, 107 (1990); J. Phys. A \textbf{24}, L175 (1991); \textit{Quantum Theory: Concepts and Methods} (Kluwer, Dordrecht, 1993), pp. 196-201.
\bibitem{cabello1} A. Cabello, and G. Garcia-Alcaine, J. Phys. A \textbf{29}, 1025 (1996); A. Cabello, J. M. Estebaranz, and G. Garcia-Alcaine, Phys. Lett. A \textbf{212}, 183 (1996); Phys. Lett. A \textbf{218}, 115 (1996).
\bibitem{penrose} R. Penrose, in \emph{Quantum Reflections}, edited by J. Ellis and A. Amati (Cambridge University
Press, Cambridge, 1994); M. Kernaghan, J. Phys. A \textbf{27}, L829 (1994); M. Kernaghan, and A. Peres, Phys. Lett. A \textbf{198}, 1 (1995).
\bibitem{cabello2}A. Cabello, and G. Garcia-Alcaine, Phys. Rev. Lett. \textbf{80}, 1797 (1998).
\bibitem{home}S. Basu, S. Bandyopadhyay, G. Kar, and D. Home, Phys. Lett. A, \textbf{279}, 281 (2001).
\bibitem{cabello4}A. Cabello, S. Filipp, H. Rauch, and Y. Hasegawa, Phys. Rev. Lett. \textbf{100}, 130404 (2008).
\bibitem{cabellosi} A. Cabello, Phys. Rev. Lett. {\bf 101} 210401(2008).
\bibitem{pitowski} P. Badzikag, I. Bengtsson, A. Cabello, and I. Pitowsky, Phys. Rev. Lett. {\bf 103} 050401(2009).
\bibitem{liu} B. H. Liu \emph{et al.}, Phys. Rev. A, {\bf 80} 044101(2009).
\bibitem{ams}E. Amselem, M. Radmark, M. Bourennane and A. Cabello, Phys. Rev. Lett. {\bf 103}, 160405 (2009).


 

\end{thebibliography}
\end{document}